\documentclass[12pt]{iopart}
\usepackage{graphicx}
\usepackage{cite}
\usepackage{amsfonts}
\usepackage{array}
\begin{document}
\title[Long-range epidemic spreading]
      {Long-range epidemic spreading with immunization}
\author{Florian Linder$^{1}$, Johannes Tran-Gia$^{1}$, Silvio R. Dahmen$^{1,2}$, and Haye Hinrichsen$^1$}
\address{$^1$ Universit\"at W\"urzburg,
	 Fakult\"at f\"ur Physik und Astronomie,\\
         D-97074 W\"urzburg, Germany}
\address{$^2$ Instituto de F{\'{\i}}sica, Universidade Federal do Rio Grande do Sul\\
	 91501-970 Porto Alegre RS, Brazil}

\date{\today}


\begin{abstract}
\noindent
We study the phase transition between survival and extinction in an epidemic process with long-range interactions and immunization. This model can be viewed as the well-known general epidemic process (GEP) in which nearest-neighbor interactions are replaced by Levy flights over distances $r$ which are distributed as $P(r)\sim r^{-d-\sigma}$. By extensive numerical simulations we confirm previous field-theoretical results obtained by Janssen {\it et al.} [Eur. Phys. J. B {\bf 7}, 137 (1999)].
\end{abstract}

\def\xvec{{\vec x}}
\def\svec{{\vec s}}
\def\kvec{{\vec k}}
\def\d{{\rm d}}

\parskip 2mm

\section{Introduction}

In statistical physics models for epidemic spreading continue to attract attention because they exhibit non-equilibrium phase transitions with universal properties. The best known example is the universality class of directed percolation (DP)~\cite{Kinzel85,Hinrichsen00,Odor04,Lubeck04}, which is observed in lattice models for epidemic spreading with nearest-neighbor infection and spontaneous recovery. Very recently, this type of phase transition was observed experimentally for the first time by Takeuchi {\it et al.}~\cite{TakeuchiEtAl07}.

In the past two decades various generalized models of epidemic spreading have been proposed. One possible generalization is to introduce a local memory in order to mimic the effect of immunization~\cite{PerelsonWeisbuch97}. This leads to the so-called general epidemic process (GEP), which has been studied both numerically and by field-theoretic methods~\cite{Cardy83a,CardyGrassberger85,Janssen85}. Another possibility is to consider long-range infections which are usually modelled by so-called Levy flights (for a recent review see~\cite{Hinrichsen07}). In such models the infection is transported over randomly chosen distances $r$, which are power-law distributed as
\begin{equation}
\label{LevyDistribution}
P(r) \sim r^{-d-\sigma}\,,
\end{equation}
where $d$ is the spatial dimension and $\sigma>0$ is an exponent controlling the characteristic shape of the distribution. By studying DP with Levy flights it turns out that critical exponents vary continuously in a certain range of $\sigma$.

Let us briefly recall some of the main results for the GEP. Starting point is the Langevin equation for directed percolation, i.e., epidemic spreading without immunization.

\begin{equation}
\partial_t \phi(\xvec,t) \;=\; a\phi(\xvec,t)-b\phi^2(\xvec,t)+D\nabla^2\phi(\xvec,t)+\xi(\xvec,t).
\end{equation}
Here $\phi$ is the density of infected individuals, $a$ is a parameter controlling the spreading rate, $-b\phi^2$ reflects to lowest order the constraint that individuals cannot be doubly infected, and $D\nabla^2\phi$ is a diffusion term accounting for nearest-neighbor interactions. The noise $\xi$ describes density fluctuations caused by the stochastic nature of the spreading process and is defined by the density-dependent correlations
\begin{equation}
\langle \phi(\xvec,t)\phi(\xvec',t')\rangle\;=\;\Gamma \,\phi(\xvec,t)\,\delta^d(\xvec-\xvec')\delta(t-t')\,.
\end{equation}
As first suggested by Cardy~\cite{Cardy83a}, the effect of immunization can be implemented by adding a term of the form 
\begin{equation}
\lambda\, \phi(\xvec,t)\,\exp\left(-w\int_0^t \d t'\, \phi(\xvec,t')\right)\,.
\end{equation}
In this term the integral sums up the past activity at position $\xvec$ between the initial condition $t=0$ and the actual time $t$. If this integrated activity is still small the exponential function is essentially equal to 1, marking a non-immune individual. However, when the integrated activity exceeds a certain threshold of the order $1/w$, the exponential function `switches' to zero, representing an immunized individual. Omitting arguments $\xvec,t$ the modified Langevin equation reads 
\begin{equation}
\partial_t \phi \;=\; a\phi-b\phi^2+D\nabla^2\phi+\xi+\lambda \,\phi\,\exp\left(w\int_0^t \d t'\, \phi(\xvec,t')\right)\,.
\end{equation}
As can be seen, the exponential function is coupled to the field $\phi(\xvec,t)$ so that it effectively modifies the spreading rate~$a$. This means that non-immune sites are infected with rate $a+\lambda$, while immune sites are infected with rate $a$. This allows one to control the rates for the first and all subsequent infections of an individual separately. The additional term modifies the critical behavior of the transition, leading to a universality class which is different from DP. This so-called GEP-class comprises all models which are defined in the spirit of this Langevin equation.

Using the Janssen-de Dominics formalism by introducing a response field $\tilde\phi(\xvec,t)$ and integrating out the Gaussian noise, one is led to a field-theoretic action~\cite{Cardy83a,CardyGrassberger85}
\begin{eqnarray}
\label{GEPAction}
S &=& \int \d^dx \int\d t \Biggl[
\tilde\phi\biggl(-\partial_t+a+D\nabla^2\biggr)\phi 
+ \frac{\Gamma}{2}\tilde\phi^2\phi - \frac{b}{2}\tilde\phi\phi^2 \\ \nonumber
&&\hspace{24mm}+
\frac{\lambda}{2}\,\tilde\phi\phi\,\exp\left(\int_0^t \d t'\, \phi(\xvec,t')\right)\Biggr]
\end{eqnarray}
which is expected to be valid in arbitrary dimensions.

Apart from the interpretation as an epidemic process with immunization, the GEP also plays an important role as a dynamical process that produces isotropic (undirected) percolation clusters in $d$ dimensions. More specifically, whenever the process terminates, it leaves behind a certain cluster of immune sites. This cluster can be shown to be an ordinary percolation cluster~\cite{StaufferAharony92} for which the critical exponents are already known (in two dimensions even exactly) and can be expressed in terms of two standard exponents $\beta$ and $\nu\equiv\nu_\perp$.  In the GEP, time as an additional degree of freedom induces another exponent, the so-called dynamical exponent $z$. Knowing $\beta$, $\nu$, and $z$, the exponents for the survival probability $P(t)\sim t^{-\delta}$ and the average number of particles $N(t)\sim t^{\theta}$ can be expressed as
\begin{equation}
\label{DeltaThetaGEP}
\delta = \frac{\beta}{\nu_\parallel} \,, \qquad \theta = \frac{d}{z} - \frac{2\beta}{\nu_\parallel}-1
\qquad \qquad \mbox{for ordinary GEP.}
\end{equation}
wehre $\nu_\parallel=z\nu_\perp$. The upper critical dimension of the GEP is $d_c=6$ and the exponents are summarized in Table~\ref{TableExponents}.

\begin{table}
\begin{center}
\begin{tabular}{c|c|c|c|c|c}
Exponent & $d=2$~\cite{MunozEtAl99}  & $d=3$~\cite{MunozEtAl99} & $d=4$~\cite{BallesterosEtAl97} & $d=5$~\cite{BallesterosEtAl97} & $d\geq 6$ \\ \hline
$\beta$ 	& 5/36  & 0.417 & 0.64 & 0.84 & 1 \\ 
$\nu_\perp$ 	& 4/3   & 0.875 & 0.68 & 0.57 & 1/2 \\ 
$\nu_\parallel$ & 1.506 & 1.169 &   &   & 1 \\
$\delta$	& 0.092 & 0.356 &   &   & 1 \\
$\theta$	& 0.586 & 0.536 &   &   & 0 \\
$z$	 	& 1.129 & 1.336 &   &   & 2 
\end{tabular}
\label{TableExponents}
\caption{Critical exponents of the standard general epidemic process. Empty entries are not available.}
\end{center}
\end{table}

In this paper we consider the GEP generalized in such a way that the nearest-neighbor interactions are replaced by non-local ones according to a Levy distribution~(\ref{LevyDistribution}). For large values of $\sigma$, where the Levy flights become short-ranged, one expects to recover the standard GEP. On the other hand, for sufficiently small values of $\sigma$ the interactions are so long-ranged that any spatial structure is wiped out, hence one expects mean-field behavior. In between there is an intermediate regime described by a non-trivial field theory~\cite{JanssenEtAl99}, where critical exponents vary continuously with $\sigma$.

\begin{figure}
\begin{center}
\includegraphics[width=160mm]{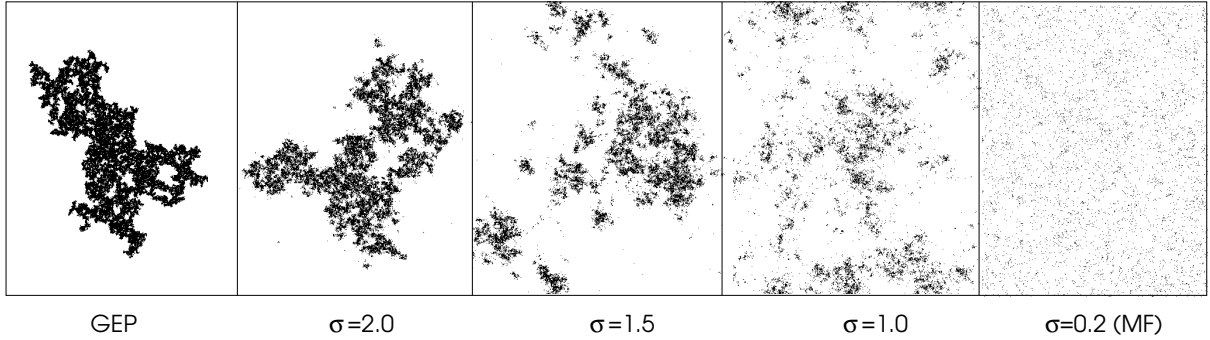}
\caption{Typical snapshot of clusters of immune sites for different values of $\sigma$.}
\label{fig:demo}
\end{center}
\end{figure}

The paper is organized as follows. In the next section we recall how the problem can be formulated as a field-theoretic action that involves a fractional derivative. Then in Sect.~\ref{AnalysisSection} we analyze the scaling behavior of this action, determine the range where perturbative methods can be applied, discuss two scaling relations which are conjectured to hold to all orders of perturbation theory, and quote the one-loop results derived by Janssen {\it et al.}~\cite{JanssenEtAl99}. Finally we verify these findings by numerical simulations in Sect.~\ref{NumericalSection}.

\section{Field-theoretic action}

A L{\'e}vy flight in $d$ dimensions is an isotropic random displacement $\xvec\to\xvec+\svec$ according to a probability distribution which for large distances decays algebraically as
\begin{equation}
\label{eq:SpatialLevyDistribution}
P(\svec) \sim |\svec\,|^{-d-\sigma}\,.
\end{equation}
Here the exponent $\sigma>0$ is a control parameter which determines the asymptotic power-law characteristics of long-range flights. In order to normalize this distribution, a lower cutoff of this power law for small $\svec$ is needed, for example in form of a minimal flight distance.

Acting on a density field $\phi(\xvec,t)$, a Levy flight generates the time evolution
\begin{equation}
\dot\phi(\xvec,t) \;\propto\;
\int {\rm d}^ds \, P(\svec)\, \biggl[\phi(\xvec+\svec,t)-\phi(\xvec,t)\biggr]\,.
\end{equation}
Introducing the non-local linear operator $\tilde{\nabla}^{\sigma}$ which acts on a function $f(\xvec)$ as
\begin{equation}
\label{eq:SpatialLevyOperator}
\tilde{\nabla}^\sigma \,f(\xvec) = \frac{1}{\mathcal{N}_\perp(\sigma)}
\int {\rm d}^ds \, |\svec\,|^{-d-\sigma} \biggl[f(\xvec+\svec)-f(\xvec)\biggr]\,,
\end{equation}
with the $\sigma$-dependent normalization constant
\begin{equation}
\mathcal{N}_\perp(\sigma)=
-\frac{\pi^{d/2}\Gamma(-\frac{\sigma}{2})}{2^\sigma \Gamma(\frac{d+\sigma}{2})}
\end{equation}
this equation of motion can be written to leading order as
\begin{equation}
\label{eq:EquationOfMotion}
\dot\phi(\xvec,t) \;=\;\tilde{\nabla}^\sigma \,\phi(\xvec,t)
\end{equation}
which has the same structure as the ordinary diffusion equation apart from the fact that the Laplacian is replaced by $\tilde{\nabla}^\sigma$. In the literature the operator $\tilde{\nabla}^\sigma$ is known as a fractional derivative because it has certain algebraic properties that generalize those of ordinary derivatives. For example, for $0<\sigma<2$ the action of the operator $\tilde{\nabla}^\sigma$ on a plane wave amounts to bringing down a prefactor of the form
\begin{equation}
\tilde{\nabla}^\sigma \, e^{i \kvec \cdot \xvec} \;=\; -|\kvec|^\sigma \, e^{i \kvec \cdot \xvec} \,.
\end{equation}
As a field-theoretic approach to GEP with Levy flights, one may simply replace the Laplacian in the action~(\ref{GEPAction}) by a fractional derivative. However, past experience in the study of directed percolation with Levy flights~\cite{JanssenEtAl99,HinrichsenHoward99,AdamekEtAl05,JimenezDalmaroni06} shows that
the fractional derivative should be added as a new term without discarding the Laplacian. The reason is that even if the Laplacian was not included in the bare action, it would be generated under renormalization group. Therefore, the field-theoretic action describing the long-ranged GEP is given by
\begin{eqnarray}
\label{GEPActionWithoutLaplacian}
S_\sigma &=& \int \d^dx \int\d t \Biggl[
\tilde\phi\biggl(-\partial_t+a+D\nabla^2+\tilde{D}\nabla^\sigma\biggr)\phi 
+ \frac{\Gamma}{2}\tilde\phi^2\phi - \frac{b}{2}\tilde\phi\phi^2 \\ \nonumber
&&\hspace{24mm}+
\frac{\lambda}{2}\,\tilde\phi\phi\,\exp\left(\int_0^t \d t'\, \phi(\xvec,t')\right)\Biggr]\,.
\end{eqnarray}
%

\section{Scaling analysis}
\label{AnalysisSection}

\subsection{Scaling dimensions}

The major problem technical problem when investigating this action lies in the structure of the exponential function. In order to apply ordinary field-theoretic methods, one needs to expand the exponential function as a power series. However, such an expansion is only meaningful if the relevance of the terms under renormalization group decrease with increasing order. 

As usual one introduces the dimensions
\begin{equation}
[\xvec]=[k]^{-1}\,,\quad
[t]=[k]^{-z}\,,\quad
[\phi]=[k]^{\chi}\,,\quad
[\tilde\phi]=[k]^{\tilde\chi}
\end{equation}
with respect to a momentum scale $k$, where $z=\nu_\parallel/\nu_\perp$ is the dynamical exponent and $\chi,\tilde\chi$ are the field exponents. As usual, the fields $\phi$ and $\tilde \phi$ stand for the creation and removal of active sites. Since each removed particle leaves an immune site behind, the scaling properties of the generated cluster of immune sites are described by the exponent~$\tilde \chi$. Therefore, one can identify $\tilde \chi$ with the exponent $\beta$ of isotropic percolation through
\begin{equation}
\tilde\chi=\beta/\nu_\perp\,.
\end{equation}
%

\subsection{Expansion of the exponential function}

Obviously, the integrand in the last term of the action has the dimension
\begin{equation}
\left[\int\d t'\, \phi(\xvec,t')\right] = [k]^{\chi-z}.
\end{equation}
This means that the exponential function can be expanded as a power series only if $\chi>z$ since in this case the terms of the series are decreasingly relevant. Under this condition the zeroth term of the series, $\frac{\lambda}{2}\tilde\phi\phi$, can be absorbed by a shift $a \to a+\lambda/2$ so that the non-trivial term to be retained in the action is the first-order term of the exponential $\frac{\lambda}{2}\tilde\phi\phi\int_0^t \d t'\, \phi(\xvec,t')$ which has the dimension $[k]^{\tilde\chi+2\chi-z}$. For $z>0$ this term is more relevant than the cubic term $-\frac{b}{2}\tilde\phi\phi^2$ in Eq.~(\ref{GEPAction}) so that the latter one can be discarded. Hence for $\chi>z>0$ the action takes the form
\begin{eqnarray}
\label{GEPAction2}
S_\sigma &=& \int \d^dx \int\d t \Biggl[
\tilde\phi\biggl(-\partial_t+a+D\nabla^2+\tilde{D}\nabla^\sigma\biggr)\phi 
+ \frac{\Gamma}{2}\tilde\phi^2\phi  \\ \nonumber
&&\hspace{24mm}+
\frac{\lambda w}{2}\,\tilde\phi\phi\,\int_0^t \d t'\, \phi(\xvec,t')\Biggr].
\end{eqnarray}
For $\sigma=2$ this expression reduces to the action of the conventional GEP.

By dimensional analysis of the field-theoretic action~(\ref{GEPAction2}) one obtains the mean field (tree level) exponents
\begin{equation}
\beta^{\rm\scriptscriptstyle MF } = 1\,,\quad
\nu_\perp^{\rm\scriptscriptstyle MF } = 1/\sigma\,,\quad
\nu_\parallel^{\rm\scriptscriptstyle MF } = 1\,,\quad
\chi^{\rm\scriptscriptstyle MF } = 2\sigma\,,\quad
\tilde \chi^{\rm\scriptscriptstyle MF } =\sigma.
\end{equation}
which are expected to be valid above the upper critical dimension
\begin{equation}
d_c=3\sigma\,.
\end{equation}
%
\subsection{Exact scaling relations}
%
The exponents $\chi,\tilde\chi$ and $z$ are related by two exact scaling relations~\cite{JanssenEtAl99}. These relations can be explained as follows.

The first scaling relation is caused by a rapidity reversal symmetry of the action. As can be verified easily, the action of the standard GEP is invariant under the replacement~\cite{Cardy83a,Janssen85}
\begin{equation}
\frac{\partial}{\partial_t}\tilde\phi(\xvec,t) \;\leftrightarrow\; \phi(\xvec,-t)\,.
\end{equation}
Obviously this rapidity reversal symmetry holds also in the case of long-range interactions $\sigma\neq 2$. Moreover, it is an exact symmetry which holds to all orders of perturbation theory. Comparing the scaling dimensions on both sides this immediately implies the exact scaling relation\footnote{Note that this scaling relation differs from the one for directed percolation, where the well-known rapidity reversal symmetry $\phi(\xvec,t)\leftrightarrow\tilde\phi(\xvec,-t)$ forces $\tilde\chi$ and $\chi$ to be identical.}
\begin{equation}
\label{FirstScalingRelation}
\tilde \chi=\chi-z.
\end{equation}
As long as the field exponent $\tilde \chi$ is positive (which is indeed the case here), the condition $\chi>z$ is fulfilled, justifying the expansion of the exponential term in Eq.~(\ref{GEPAction2}).

The second scaling relation can be made plausible as follows. The term with Levy operator $\tilde{D}\tilde\phi\nabla^\sigma\phi$ is bilinear in the fields. Therefore, it modifies the free theory, whereas it does not change the interactions described by the cubic terms. In other words, the Levy term modifies the free propagator of the theory, but it has no influence on the structure of the Feynman diagrams. This means that all loop integrals have exactly the same form as in the standard GEP, the only difference being that the free propagator in momentum space $G^{-1}=a+i\omega+D k^2$ has to be replaced by $G^{-1}=a+i\omega+D k^2+\tilde D |k|^\sigma$. In so far the situation is completely analogous to the case of directed percolation with long-range interactions.

In the case of Levy-DP it was observed that the corrections caused by the loop integrals can be expressed as a power series in the momentum $k$. As in any Taylor expansion, this series involves only integral powers, the leading order being $k^2$. This means that the long-ranged Levy operator produces loop corrections that can be absorbed in the coefficient of the (short-range) Laplacian. In other words, the Levy term is not renormalized, but instead renormalizes the short-range diffusion coefficient $D$. This implies that the Levy term with scaling dimension $[k]^{\chi+\tilde\chi+\sigma-d-z}$ must be invariant under rescaling, hence one obtains the scaling relation
\begin{equation}
\label{Hyperscaling}
\chi+\tilde\chi+\sigma-d-z=0
\end{equation}
This so-called hyperscaling relation holds to all orders of perturbation theory below the upper critical dimension $d\leq d_c$. Combining it with Eq.~(\ref{FirstScalingRelation}) one obtains
\begin{equation}
2 \tilde\chi = d-\sigma  \,,
\end{equation}
or equivalently
\begin{equation}
\beta = \frac{\nu_\perp}{2}(d-\sigma)\,.
\end{equation}
The hyperscaling relation implies that the exponents
\begin{equation}
\delta=\frac{\beta}{\nu_\parallel} \,, \qquad \theta=\frac{d \nu_\perp-2\beta}{\nu_\parallel}-1
\end{equation}
are related by
\begin{equation}
\frac{\delta}{\theta+1}\;=\; \frac{d-\sigma}{2\sigma}\,.
\end{equation}
%

%
\subsection{Crossover from long-ranged to ordinary GEP}
%
The scaling relations are particularly interesting at the crossover from the regime, where the Levy operator is relevant, to the ordinary GEP, where the short-ranged Laplacian plays the dominant role. This crossover takes place at a particular value of $\sigma$ denoted as $\sigma_s$. Assuming that the crossover is smooth in the sense that the exponents do not change discontinuously, $\sigma_s$ can be computed by plugging the known GEP exponents into the hyperscaling relation~(\ref{Hyperscaling}) and solving it for $\sigma$:
\begin{equation}
\sigma_s \;=\; d - \frac{2\beta^{\scriptscriptstyle\rm GEP}}{\nu_\perp^{\scriptscriptstyle\rm GEP}}.
\end{equation}
For example, in $d=2$ dimensions the threshold for $\sigma$, above which the short-range GEP is recovered, is given by
\begin{equation}
\sigma_s = \frac{43}{24} \approx 1.792 \qquad \qquad (d=2)\,.
\end{equation}
At a first glance this result seems to be counterintuitive as it contradicts the naive expectation that the fractional derivative $\tilde{\nabla}^\sigma$ should reduce to the ordinary Laplacian $\nabla^2$ at the threshold $\sigma=2$. However, studies of directed percolation and the Ising model with Levy-type interactions have shown that this naive argument does not necessarily hold in an interacting theory. Moreover, it is surprising that in 2D the crossover takes place at a threshold $\sigma_s< 2$, while in previous studies of directed percolation $\sigma_s$ was always found to be larger than~2.

\subsection{One-loop results}
%
According to Janssen {\it et al.}~\cite{JanssenEtAl99}, the critical exponents computed in $d=d_c-\epsilon=3\sigma-\epsilon$ dimensions to one-loop order are given by 
\begin{eqnarray}
\beta  &=& 1-\frac{\epsilon}{4\sigma}+\mathcal{O}(\epsilon^2)\, , \\
\nu_\perp &=& \frac{1}{\sigma} + \frac{\epsilon}{4\sigma^2}+\mathcal{O}(\epsilon^2)\,,\\
\nu_\parallel &=& 1+\frac{\epsilon}{16\sigma}+\mathcal{O}(\epsilon^2)\,,\\
z &=& s-\frac{3\epsilon}{16}+\mathcal{O}(\epsilon^2)\,,\\
\delta &=& 1-\frac{5 \epsilon}{16 \sigma}+\mathcal{O}(\epsilon^2)\,,\\
\theta &=& \frac{3 \epsilon}{16 \sigma} + \mathcal{O}(\epsilon^2)\,.
\end{eqnarray}
Usually a field-theoretic $\epsilon$-expansion can be verified only approximately by numerical simulations: since the dimension $d$ is an integer number, it is in most cases impossible to study small values of $\epsilon \ll 1$. However, in the present model the upper critical dimension $d_c=3\sigma$ is controlled by the parameter $\sigma$ and can be chosen in such a way that $\epsilon$ becomes small. This allows one to test the one-loop results directly through Monte Carlo simulations, as we show in the next section.

\section{Numerical analysis}
\label{NumericalSection}
%
\subsection{Strategy}
%
For a simulation in $d$ dimensions, let us introduce the parameter $\mu$ by
\begin{equation}
\mu \;:=\; \sigma -d/3\,.
\end{equation}
For $\mu=0$ the $d$-dimensional system is at the upper critical dimension, while for a small $\mu>0$ it is slightly below $d_c$ with $\epsilon=d_c-d=3\mu$. Substituting this parameter the exponents $\delta$ and $\theta$ are given to one-loop order by
\begin{equation}
\label{MuCorr}
\delta = 1-\frac{45 \mu}{16 d}+\mathcal{O}(\mu^2) \,, \qquad
\theta = \frac{27 \mu}{16 d}+\mathcal{O}(\mu^2)\,.
\end{equation}
Hence, plotting these exponents as functions of $\sigma$, one expects the following behavior:
\begin{itemize}
\item In the mean field regime $\sigma< d/3$ the exponents are constant and given by their mean field values. 
\item In the interval $d/3 < \sigma < \sigma_s$ the exponents vary continuously. The slope of the tangent at the left edge is just one-loop correction in Eq.~(\ref{MuCorr}).
\item Finally, for $\sigma>\sigma_s$ the Levy operator becomes irrelevant and the exponents of the standard GEP are recovered.
\end{itemize}
%


\begin{figure}
\makebox[50mm][l]{
\includegraphics[width=80mm]{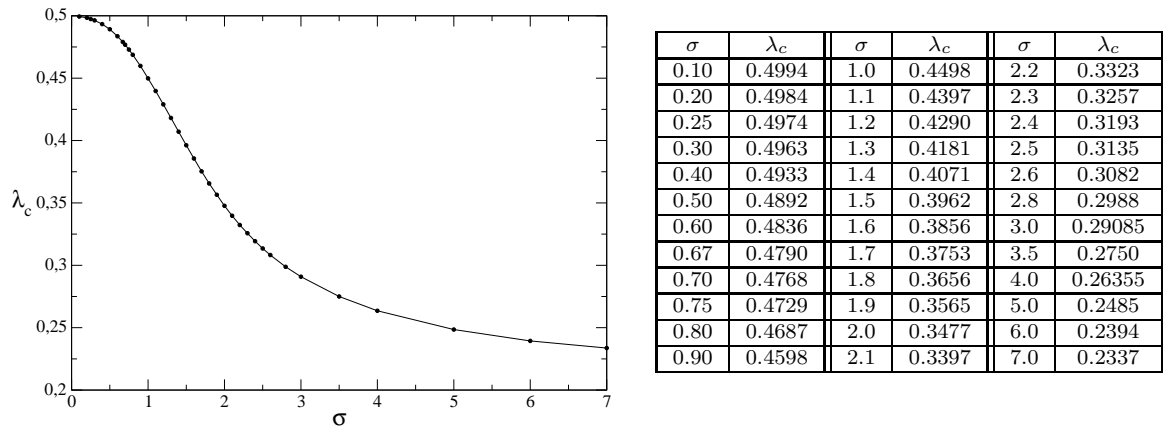}}
\vspace{-55mm}
\flushright{\makebox[50mm][r]{
\begin{scriptsize}
\begin{tabular}{|c|c||c|c||c|c|}
      \hline
      $\sigma$ & $\lambda_c$ & $\sigma$ & $\lambda_c$ & $\sigma$ & $\lambda_c$ \\
      \hline
      0.10 & 0.4994 & 1.0 & 0.4498 & 2.2 & 0.3323 \\
      \hline
      0.20 & 0.4984 & 1.1 & 0.4397 & 2.3 & 0.3257 \\
      \hline
      0.25 & 0.4974 & 1.2 & 0.4290 & 2.4 & 0.3193 \\
      \hline
      0.30 & 0.4963 & 1.3 & 0.4181 & 2.5 & 0.3135 \\
      \hline
      0.40 & 0.4933 & 1.4 & 0.4071 & 2.6 & 0.3082 \\
      \hline
      0.50 & 0.4892 & 1.5 & 0.3962 & 2.8 & 0.2988 \\
      \hline
      0.60 & 0.4836 & 1.6 & 0.3856 & 3.0 & 0.29085 \\
      \hline
      0.67 & 0.4790 & 1.7 & 0.3753 & 3.5 & 0.2750 \\
      \hline
      0.70 & 0.4768 & 1.8 & 0.3656 & 4.0 & 0.26355 \\
      \hline
      0.75 & 0.4729 & 1.9 & 0.3565 & 5.0 & 0.2485 \\
      \hline
      0.80 & 0.4687 & 2.0 & 0.3477 & 6.0 & 0.2394 \\
      \hline
      0.90 & 0.4598 & 2.1 & 0.3397 & 7.0 & 0.2337 \\
      \hline
\end{tabular}
\end{scriptsize}}}
\vspace{5mm}
\caption[Critical percolation threshold as a function of $\sigma$]{Critical percolation threshold $\lambda_c$ of the long-range general epidemic process as a function of $\sigma$ (see text).}
\label{lambdac}
\end{figure}

\subsection{Details of the simulation}
%
\noindent
The model studied in our simulations is defined on a two-dimensional square lattice of sites that can be in the states `healthy' ($H$), `contaminated' ($C$) and `immune' ($I$). The model evolves by the following microscopic transitions: On the one hand, each contaminated site becomes immune at a rate $\lambda$
\begin{center}
$C\stackrel{\lambda}{\rightarrow}I$.
\end{center}
On the other hand, a contaminated site may infect another site, which is randomly selected by a Levy flight at distance $r$:
\begin{center}
$C\stackrel{\mbox{\tiny Levy flight}}{\cdots\cdots}H\stackrel{1-\lambda}{\rightarrow}C{\cdots\cdots}C$\,.
\end{center}
These transitions were implemented by random-sequential updates on the two-dimensional grid with periodic boundary conditions and lateral length $L$. An integer variable $s(x,y)$ is attached to each lattice site $(x,y)$ and stored in a two-dimensional array. $s(x,y)=0$ means that the site is in state $H$, $s(x,y)=1$ stands for a site in state $C$ and $s(x,y)=2$ denotes a site in state $I$. In addition, we kept all infected sites in a dynamically generated list to accelerate the numerical calculations. The dynamics of the simulation is generated by a repetition of the following steps:
\begin{quote}
\begin{enumerate}
\item[1)] Randomly choose one out of the list of infected sites.\\[-3mm]
\item[2)] With the probability $\lambda$ the site is immunized by the transition $C\to I$, followed by a jump to step 4.\\[-3mm]
\item[3)] Otherwise, a second site is selected by a Levy flight according to the $\sigma$-dependent distribution~(\ref{LevyDistribution}) (see below). If this target cell is in the `healthy' state, it will be infected by the transition from state $H$ to $C$.\\[-3mm]
\item[4)] If the number of infected sites is zero, the simulation stops, otherwise increase the system time $t$ by the reciprocal total number of infected sites and go back to step 1.
\end{enumerate}
\end{quote}
Hence the model is controlled by two parameters, namely, the Levy exponent $\sigma>0$ and the critical parameter $\lambda\in[0,1]$.

Upon infection, the isotropic Levy flight used to locate the second site is distributed by a power-law $P(r)\sim r^{-2-\sigma}$ where $r=\sqrt{\Delta x^2+\Delta y^2}$ is the absolute value of the distance between two sites. In two dimensions, the radial distance $r$ and the equally distributed angle $\phi$ can be generated by setting $r:=z_1^{-1/\sigma}$ and $\phi:=2 \pi z_2$, where $z_{1,2}\in[0,1]$ are random numbers. Note that the radial distribution has a lower cutoff so that the minimal flight distance is $1$. The actual coordinates are then computed by rounding the cartesian coordinates $(\Delta x,\Delta y)=(r \cos \phi,r\sin \phi)$ to integers. 

In order to avoid the time-consuming use of trigonometric functions, one may accelerate simulations significantly by doing the following: first one repeatedly generates a random vector in a square with real-valued coordinates $\Delta x_s,\Delta y_s\in[0,1]$ until $r_s^2={\Delta x_s}^2+{\Delta y_s}^2<1$. This procedure generates an isotropically distributed random vector on the unit disk and with a squared radius $r_s^2$ equally distributed between $0$ and~$1$. Then this radial distribution has to be corrected in such a way that the desired Levy distribution~(\ref{LevyDistribution}) is obtained. This can be done by setting
\begin{eqnarray}
\Delta x &:=& {\Delta x_s \over r_s} \cdot r = \Delta x_s \cdot \left({\Delta x_s}^2+{\Delta y_s}^2\right)^{-1/2-1/{\sigma}} \nonumber \\
\Delta y &:=& {\Delta y_s \over r_s} \cdot r = \Delta y_s \cdot \left({\Delta x_s}^2+{\Delta y_s}^2\right)^{-1/2-1/{\sigma}} 
\end{eqnarray}
Levy flights exceeding the system size are treated by applying periodic boundary conditions. As discussed in Ref.~\cite{HinrichsenHoward99}, alternative methods where such flights are dismissed or truncated are plagued by much stronger finite-size effects.

\subsection{Numerical results}
%
In order to estimate the critical exponents of anomalous DP we used dynamical simulations at criticality, where a critical cluster is grown from a single active seed (just as in Fig. \ref{fig:demo}). In our simulations, the initial condition was a $L\times L$ - lattice with healthy sites, in which we placed four neighboring contaminated sites forming a seed. Averaging over many independent realizations we measured the survival probability~$P(t)$ and the actual number of contaminated sites $N(t)$. At criticality, these quantities are expected to scale as
\begin{eqnarray}
P(t) \propto t^{-\delta} \nonumber
\hspace{0.01\textwidth},
\hspace{0.01\textwidth}
N(t) \propto t^\theta \nonumber
\hspace{0.01\textwidth},
\end{eqnarray}
where $\delta=\beta /\nu_\parallel$ and $\theta=(d\nu_\perp-2\beta)/\nu_\parallel -1$. Since deviations from criticality lead to a curvature of $P(t)$ and $N(t)$ in double logarithmic plots, we held $\sigma\in[0.2,7.0]$ constant for different values of $\lambda$, searching for a straight line in order to get a precise estimate of the percolation threshold $\lambda$ for different values of $\sigma$ (see Fig. \ref{lambdac}). For small $\sigma$, finite size effects even occur at small simulation times, which can be minimized by increasing the lateral system size $L$. For $\sigma=0.2$ and $L=14000$, as an example, finite size effects occurred for $T>2000$. For bigger $\sigma$, we simulated up to $T=10000$ and averaged over $5000$ runs.

\begin{figure}
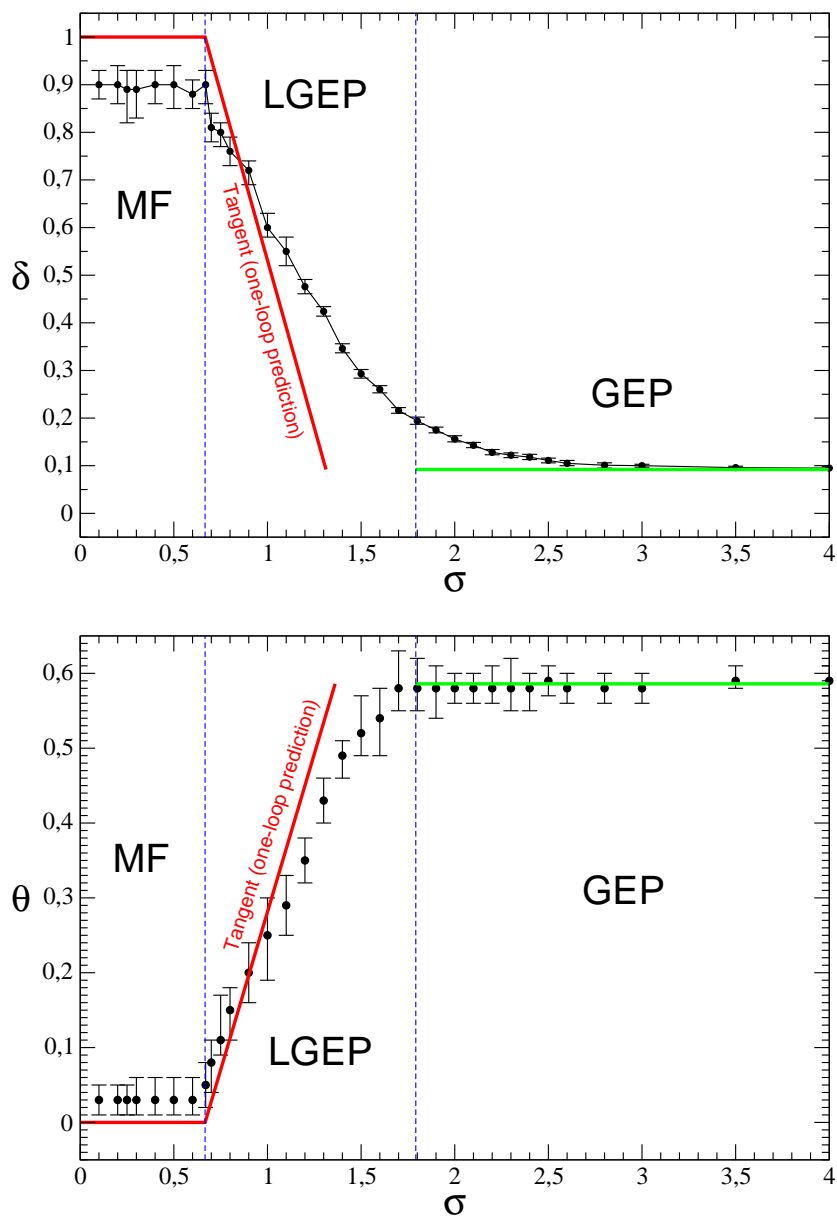

\begin{center}
\includegraphics[width=110mm]{delta}\\[5mm]
\includegraphics[width=110mm]{theta}
\caption{Exponents $\delta$ and $\theta$ for epidemic spreading with immunization as a function of the Levy exponent $\sigma$ in the mean field (MF), long-range (LGEP) and short-range regime (GEP).}
\label{fig:deltatheta}
\end{center}
\end{figure}

As can be seen from Fig.~\ref{fig:deltatheta} one observes three different regimes as expected (see discussion after equation~(\ref{MuCorr})):
\begin{itemize}
 \item $\sigma < 2/3$: mean-field behavior (interactions range is effectively infinite).
\item $2/3 < \sigma <\sigma_{s}$: intermediate phase with continuously varying exponents whose values, to one-loop approximation, are given by~(\ref{MuCorr}).
\item $\sigma >\sigma_{s}$: ordinary short-range GEP with exponents given in Tab.~\ref{TableExponents}, where the Levy Operator becomes irrelevant.
\end{itemize}
Our numerical results are shown in Fig.~\ref{fig:deltatheta}, where the expected values of the mean-field exponents and the short-range GEP exponents are marked by horizontal lines. The field-theoretic one-loop predictions according to~(\ref{MuCorr}) are shown as tilted straight lines, which in principle should be tagent to the numerical curve near the crossover. The coincidence is far from being perfect, probably because of strong finite-size effects when $\sigma$ is small. Nevertheless the estimates seem to follow the predicted behavior and, as far as we can jugde, confirm the field-theoretic results.

\section{Conclusions}

In this paper we have studied the general epidemic process (GEP) with long-range interactions and immunization. In our model infections are transported by L{\'e}vy flights over large distances according to a power-law distribution $P(r)\sim f^{-d-\sigma}$ depending on a parameter $\sigma>0$. After being infected once, individuals become perfectly immune.

The main result of our work is that we fully confirm, through numerical analysis, the field-theoretic predictions of Janssen {\it et al.}, namely that one should observe three different regimes depending on the value of $\sigma$. For $\sigma<2/3$, the system is so strongly mixed that mean-field behavior is expected at the transition. On the other hand, for $\sigma$ above a characteristic threshold $\sigma_s\approx 1.792$ the usual short-range Laplacian $\nabla^2$ dominates over the Levy operator $\nabla^{\sigma}$ which, in field-theoretic parlance, becomes irrelevant. In this regime the system becomes effectively short-ranged and one recovers GEP exponents. In between, one has an intermediate regime with continuously varying scaling exponents. To one-loop order these exponents are given by~(\ref{MuCorr}), in fair agreement with the numerical simulations for small values of $\sigma>2/3$.

The $\sigma$-dependent simulation is interesting in that it allows us to move continuously from the mean-field regime over the long-range phase to the short-range regime without changing the space dimension $d$. At the crossover from the mean field to the LGEP phase, it is possible to verify the field theoretic one-loop prediction quantitatively. The second crossover from the LGEP phase to the short-range GEP regime is also interesting in so far as it takes place, for $d=2$, at $\sigma_s\approx 1.792$ and not at $\sigma=2$, as one would naively expect as a  threshold where $\nabla^{\sigma}$ replaces the ordinary Laplacian $\nabla^2$. \\

\noindent\textbf{Acknowledgements:}\\
S. R. Dahmen would like to acknowledge for financial support the
Alexander-von-Humboldt Foundation and the TP3 Group in the
University of W\"urzburg for their hospitality.


\vspace{4mm}
\noindent{\bf References:}

\begin{thebibliography}{10}
\expandafter\ifx\csname url\endcsname\relax
  \def\url#1{{\tt #1}}\fi
\expandafter\ifx\csname urlprefix\endcsname\relax\def\urlprefix{URL }\fi
\providecommand{\eprint}[2][]{\url{#2}}

\bibitem{Kinzel85}
Kinzel W 1985 {\em Z. Phys. B\/} {\bf 58} 229

\bibitem{Hinrichsen00}
Hinrichsen H 2000 {\em Adv. Phys.\/} {\bf 49} 815 [cond-mat/0001070]

\bibitem{Odor04}
\'Odor G 2004 {\em Rev. Mod. Phys.\/} {\bf 76} 663

\bibitem{Lubeck04}
L{\"u}beck S 2004 {\em Int. J. Mod. Phys. B\/} {\bf 18} 3977

\bibitem{TakeuchiEtAl07}
Takeuchi K~A, Kuroda M, Chaté H and Sano M 2007 {\em Phys. Rev. Lett.\/} {\bf
  99} 234503

\bibitem{PerelsonWeisbuch97}
Weisbuch A~S~P~G 1997 {\em Rev. Mod. Phys.\/} {\bf 69} 1219

\bibitem{Cardy83a}
Cardy J~L 1983 {\em J. Phys. A: Math. Gen.\/} {\bf 16} L709

\bibitem{CardyGrassberger85}
Cardy J~L and Grassberger P 1985 {\em J. Phys. A: Math. Gen.\/} {\bf 18} L267

\bibitem{Janssen85}
Janssen H~K 1985 {\em Z. Phys. B\/} {\bf B 58} 311

\bibitem{Hinrichsen07}
Hinrichsen H 2007 {\em J. Stat. Mech.\/}  P07066

\bibitem{StaufferAharony92}
Stauffer D and Aharony A 1992 {\em Introduction to Percolation Theory\/}
  (London: Taylor \& Francis)

\bibitem{MunozEtAl99}
Mu{\~n}oz M~A, Dickman R, Vespignani A and Zapperi S 1999 {\em Phys. Rev. E\/}
  {\bf 59} 6175

\bibitem{BallesterosEtAl97}
Ballesteros H, Fernandez L, Martin-Mayor V, Sudupe A, Parisi G and Ruiz-Lorenzo
  J 1997 {\em Phys. Lett. B\/} {\bf 400} 346

\bibitem{JanssenEtAl99}
Janssen H~K, Oerding K, van Wijland F and Hilhorst H~J 1999 {\em Eur. Phys. J.
  B\/} {\bf 7} 137

\bibitem{HinrichsenHoward99}
Hinrichsen H and Howard M 1999 {\em Eur. Phys. J. B\/} {\bf 7} 635 643

\bibitem{AdamekEtAl05}
Adamek J, Keller M, Senftleben A and Hinrichsen H 2005 {\em J. Stat. Mech.:
  Theor. Exp.\/}  P09002

\bibitem{JimenezDalmaroni06}
Jimenez-Dalmaroni A 2006 {\em Phys. Rev. E\/} {\bf 75} 011123

\end{thebibliography}

\providecommand{\newblock}{}

\end{document}